\documentclass[aps,prb]{revtex4}
\usepackage{graphicx,amssymb,amsfonts,amsmath}

\begin{document}

\title{Coulomb Gas on the Keldysh Contour: Anderson-Yuval-Hamann representation
of the Nonequilibrium Two Level System
}
\author{Aditi Mitra}
\affiliation{Department of Physics, New York University, 4
Washington Place, New York, NY 10003}
\author{A. J. Millis}
\affiliation{\\Department of Physics, Columbia University, 538 W.
120th St., New York, NY 10027}
\date{\today}


\begin{abstract}
The nonequilibrium tunneling center model of   a localized electronic
level coupled to a fluctuating two-state system and to two
electronic reservoirs, is solved via an Anderson-Yuval-Hamann
mapping onto a plasma of alternating positive and negative charges
time-ordered along the two ``Keldysh" contours needed to describe
nonequilibrium physics. The interaction between charges depends
both on
whether their time separation is small or large compared to a
dephasing scale defined in terms of the chemical potential
difference between the electronic reservoirs and
on whether their time separation is larger or smaller than
a decoherence scale defined in terms of the current flowing from one
reservoir to another. A renormalization group transformation
appropriate to the nonequilibrium problem is defined. An important
feature is the presence in the model of a new coupling, essentially
the decoherence rate, which acquires an additive renormalization
similar to that of the energy in equilibrium problems. The method is
used to study interplay between the dephasing-induced formation of
independent resonances tied to the two chemical potentials and the
decoherence which cuts off the scaling and leads to effectively
classical long-time behavior. We determine the effect of departures
from equilibrium on the localization-delocalization phase
transition. 

\end{abstract}


\maketitle

\section{Introduction}

Understanding the nonequilibrium behavior of interacting quantum
mechanical systems is one of the important open issues in condensed
matter physics, with applications in nanoscience~\cite{Paaske06},
the study of cold atoms in optical lattices \cite{Morsch06},
nonlinear spectroscopies~\cite{Axt98} and, transport at quantum
critical points~\cite{Hogan06}. One may distinguish three classes of
nonequilibrium situations: response of a system initially in  an
equilibrium state to a strong transient pulse, time evolution of a
system from a particular initial condition, and the steady state
behavior of a driven system. In this paper we shall be concerned
with one of the simplest examples of the third class of problems: a
quantum mechanical system with only a few degrees of freedom,
coupled to two reservoirs with which particles and energy may be
exchanged, and with the nonequilibrium drive arising from a
difference, $\Delta \mu$, in chemical potential $\mu$ between the
reservoirs. This model is of experimental relevance in the context of
single molecule devices~\cite{singlemolecule} and of quantum 
dots~\cite{kondoinquantumdot} and is important
as a paradigm problem for the development of techniques and insights. 
Two crucial issues in nonequilibrium physics are dephasing and decoherence. 
In the model we study dephasing arises because the wave functions in the two reservoirs evolve 
in time at rates which differ by $\Delta \mu$ whereas decoherence 
arises from the flow of energy and particles across the system. 
An important issue in nonequilibrium physics is to develop methods
which allow these effects to be systematically analyzed.

On the formal level, investigation of equilibrium systems is based
on the partition function. Powerful techniques, most importantly the
renormalization group method, enable one to eliminate putatively
unimportant degrees of freedom and derive an effective theory governing
the low energy behavior of interest. The renormalization group
has been implemented in two intimately related ways: by considering
changes in the self energies, vertex functions, and correlation
functions in diagrammatic calculations, and by working
directly with the partition function,  generating an effective action
describing only the degrees of freedom of interest. 

Most applications of renormalization group ideas to nonequilibrium
problems have been based on the first approach: a diagrammatics
is constructed using the Keldysh technique and then the flow of
vertices, self energies and response functions under changes
in cutoff is studied.  In pioneering work  Rosch and co-workers~\cite{Rosch01,Rosch03} 
constructed a nonequilibrium scaling theory 
for the Kondo problem by identifying logarithms in perturbative
calculations. Measurable quantities are computed, scaling equations for
the coupling constants of the Hamiltonian were inferred from
logarithmic dependences of observables on the upper frequency
cutoff.  In a very recent paper Borda et al \cite{Borda07}
used similar techniques to study the nonequilibrium behavior
of the tunneling center model by perturbation theory in the dot-lead coupling.
Paaske et. al provided further insight
\cite{Paaske04a,Paaske04b} into the physically crucial issue
of  decoherence rates, showing that voltage-induced decoherence
enters differently into different observables,
so that  the analogy between temperature and
decoherence is not precise. In a more recent set of approaches, a
transformation is performed on the Hamiltonian itself.~
\cite{Kehrein05,Gezzi07} 

While these approaches have established
a number of basic results and concepts, the development of the subject
remains incomplete. In equilibrium, defining a renormalization
group transformation directly on the free energy provided important
insights into the meaning of the transformations and the formal
structure of the theory. A similar analysis out of equilibrium
should lead to valuable insights including a clearer understanding of
decoherence and dissipation and a more precise definition of the
charges in the renormalization group equations
and the ability to map one problem onto another.  In this paper we
therefore examine a simple model, the tunneling center problem, 
from the effective action point of view. The ``tunneling center" 
model is perhaps the simplest
example of a wide class of quantum impurity models such as the Kondo
problem or the spin-boson problem, which  involve a
finite number of local degrees of freedom coupled to reservoirs. The
equilibrium behavior of these models is very rich, involving
nontrivial correlated states, dynamically generated energy scales
and quantum phase transitions \cite{TLS}. The equilibrium physics
revolves around a competition between formation of a quantum
coherent state of the local degrees of freedom and the decoherence
associated with coupling to the reservoir.  We wish to 
characterize the effect of departures from equilibrium
on this physics.

Effective actions for nonequilibrium problems have been discussed
by several authors; for a review see e.g. \cite{Kamenev04}. Here we 
define and analyze a renormalization group transformation directly
on the effective action. Our approach is somewhat similar
to previous work of K\"{o}nig and collaborators,~\cite{Konig96,Konig00}
who defined a renormalization group transformation directly
on the equation of motion for  the density matrix. These authors
treated the dot-lead coupling via (self-consistently resummed)
perturbation theory. We study a simpler model
in which we are able to treat the dot-lead coupling exactly.
Our formalism allows us to deal with
decoherence and orthogonality physics on the same footing. 
We show that the dephasing scale
$\Delta \mu$ defines a crossover beyond which both the physics
and the formalism changes. We find that the effective theory
valid for energy scales lower than $\Delta \mu$ is characterized by a richer
structure of charges than is the corresponding equilibrium theory. It also
involves, as a new parameter, a decoherence rate. This arises physically
from the flow of current through the system. It enters the theory as an additional
parameter, which is subject to additive renormalization (as is the total energy in 
usual RG treatments). This is discussed 
in more detail in section~\ref{Scaling}.

The rest of the paper is organized as follows. In section
~\ref{Formulation} we define the model we consider and obtain the
effective action. In section \ref{Scaling} we derive the relevant
scaling equation, and in section \ref{Solution} we present the
solution and its physical content, among other things displaying the
different physics of decoherence and dephasing. 
Section \ref{Conclusion} is a summary and conclusion, which outlines
implications for other problems.
\begin{figure}[t]
\begin{center}
\includegraphics
[width=0.3\columnwidth]{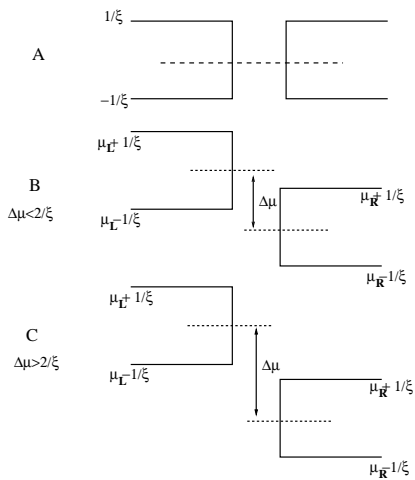}
\end{center}
\caption{Sketch of energy levels and cutoffs; band energy cutoffs
chosen for each lead to be symmetrical about Fermi energy in that
lead. (A) (upper panel): chemical potential difference $\Delta
\mu=\mu_L - \mu_R = 0$. (B) (middle panel) Chemical potential difference $\Delta \mu
\neq 0$ but less than cutoff $2\xi^{-1}$. (C) (lower panel)  Chemical
potential difference $\Delta \mu$ greater  than cutoff $2\xi^{-1}$;
no dissipation processes possible.} \label{VD}
\end{figure}

\section{Model and derivation of the Coulomb gas \label{Formulation}}

There are many realizations of the tunneling center model~\cite{SBexamples}. 
We 
consider a two-state system, which we represent in
spin notation, linearly
coupled to the density of electrons in an N-fold degenerate
electronic level (creation operator $d^{\dagger}_{\alpha}$ with
$\alpha=1...N$), which is itself hybridized with two ($L$ and $R$)
leads characterized by free-fermion statistics with possibly
different chemical potentials i.e., $\Delta\mu = \mu_L - \mu_R \neq
0 $. We consider mainly temperature $T=0$. We assume that the
coupling to the leads preserves the ``pseudospin" index $\alpha$.
The Hamiltonian may then be written as ($S_{z,x} =
\frac{1}{2}\sigma_{x,z}$ is a spin matrix)
\begin{eqnarray}
H&=& H_{loc} + H_{bath} \label{hdef}\\
H_{loc} &=& S_z B + \Delta_T S_x  +  \lambda D S_z \sum_{\alpha =
1..N}d^{\dagger}_{\alpha } d_{\alpha} \label{hloc}\\
H_{bath} &=& \sum_{a=L,R,\alpha=1..N}\int d\epsilon
\frac{\epsilon}{D} c_{\epsilon a\alpha}^{\dagger} c_{\epsilon a
\alpha} + \sqrt{\frac{1}{\pi}} \int d\epsilon \sum_{a =
L,R,\alpha=1..N} \left(
\cos\theta_{a} c_{\epsilon a \alpha}^{\dagger} d_{\alpha} + h.c.\right)\nonumber \\
\label{hb}
\end{eqnarray}
where $0\leq \theta_L = \frac{\pi}{2}-\theta_R\leq \frac{\pi}{2}$
and we have absorbed the mean hybridization strength into the
variable $\varepsilon$. The density of states per pseudospin
$\alpha$ in the leads is $D^{-1}$. For the two level system we
choose the spin basis which diagonalizes the coupling to the $d$
electrons, and parameterize this coupling by a dimensionless
variable $\lambda$ and the lead density of states. The ``magnetic
field" $B$ is the level splitting of the two level system and the
parameter $\Delta_T$ gives the tunneling between the states. Note that
if $B=0$ the Hamiltonian has a particle-hole symmetry, so its
energies are invariant under combined operations of changing the
state of the two level system from ``up" to ``down"  and changing
particles to holes ($d\leftrightarrow d^\dagger$); this simplifies
the algebra without changing the conclusions.

The model requires an upper cutoff, $\xi^{-1}$ for the  energy
integrals. It is most convenient to assume that in each lead the
cutoff is symmetrical about the chemical potential in that lead,
{\sl i.e.} in lead $a$ the energy integrals run from
$\mu_a-\xi^{-1}$ to $\mu_a+\xi^{-1}$ (see Fig.~\ref{VD}). We assume
that in the starting model the cutoff energies are much larger than
the chemical potential difference or than the level splitting
parameter $B$.

Crucial parameters of the model are the nonequilibrium phase shifts
introduced in~\cite{Ng96} and defined by
\begin{equation}
\delta_{a}= \arctan{\left[\frac{\lambda \cos^2\theta_a}{1 -\frac{\lambda^2}{4}- i
sgn(\mu_a -\mu_{\bar{a}})\lambda \sin^2\theta_a}\right]}
\label{phaseshift}
\end{equation}
Note that the phase-shifts are not independent variables, but are
related to each other as $\delta_{eq}=\delta_L +
\delta_R=\arctan\lambda/(1-\lambda^2/4) $.
Also note that while $\delta_{eq}$ is just the difference of the equilibrium phase shifts 
$\pm \tan^{-1}\frac{\lambda}{2}$
associated with the two states $S_z=\pm \frac{1}{2}$, $\delta_{L,R}$ are not the differences of the nonequilibrium
phase shifts associated with the two states.

The behavior of the two level system is specified by the reduced
density matrix, given at time $t$ in terms of an initial condition
$\rho(0)$ at time $t=0$ by
\begin{eqnarray}
\hat{\rho}_{S} = Tr_{el}\left[e^{-i H t} \rho(0) e^{i H t}\right]
\end{eqnarray}
Here $Tr_{el}$ indicates a trace over all of the electronic degrees
of freedom. We follow Anderson, Yuval and Hamann  \cite{Anderson69}
and expand $\hat{\rho}_{S}$
perturbatively in the ``spin-flip" amplitude $\Delta_T$. The spin
flip events are viewed as particles with ``fugacity" $ln \left[\xi \Delta_T\right]$ and interaction
determined by the trace over electrons. The new features are the
need for two time contours and a dependence of the interaction on
the chemical potential difference $\Delta \mu$. 

\begin{figure}[t]
\begin{center}
\includegraphics
[width=0.5\columnwidth]{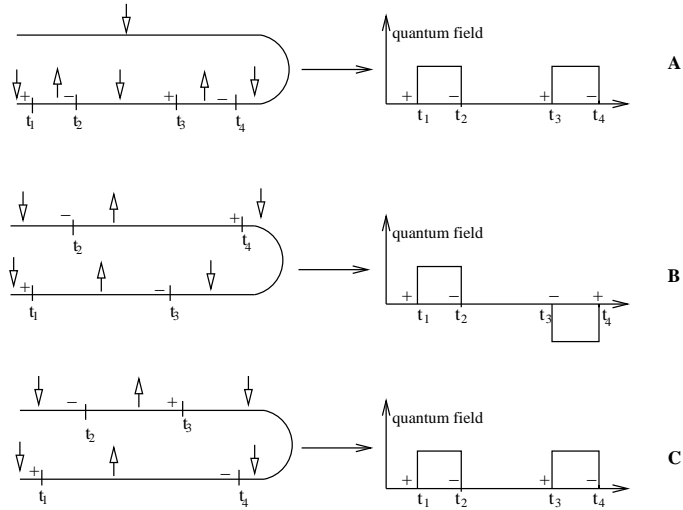}
\end{center}
\caption{ Examples  of spin-flip events in the Keldysh two axis 
representation (left side), and in the single axis representation employing 
quantum and classical fields (right side).  The charges $n_i$ are indicated on the Figure.
In the top and bottom examples the fields $q=+1$ for all instantons. In the middle example
$q=+1$ for the left hand pair of instantons and $=-1$ for the right hand pair. 
Although the first and
the third examples have the same quantum field configuration, but the different
positions of the times on the Keldysh contour leads to different phase factors. 
} \label{kdq}
\end{figure}

A term in the expansion of the density matrix consists of $n_{-}$ spin-flip events at
times running from $0_{-}$ to $t$ along the time-ordered contour,
followed by $n_{+}$ anti-time-ordered from $t$ to $0_{+}$. Some examples are shown on
the left side of
Fig \ref{kdq}: the top panel has $n_+=4$, $n_-=0$ whereas in the two lower panels
$n_+=n_-=2$. Labeling
the times in this ``Keldysh order" (time ordered on the $-$ contour
and anti-time ordered on the $+$ contour) we have for the diagonal components of the
density matrix 
($\sigma=\pm1$ below represents the state  of the
impurity spin),
\begin{eqnarray}
\langle \sigma |\hat{\rho}_S(t) |\sigma\rangle &= \sum_{n_- + n_+=
even} (-i)^{n_-} i^{n_+}
\langle \sigma | \tau_x^{n_-}\rho_{S0} \tau_x^{n_+} |\sigma \rangle \label{couleq}\\
& \left[\int_{0}^{t}\frac{dt_{n_-}}{\xi} \int_{0}^{t_{n_-}}
\frac{dt_{n_--1}}{\xi} \ldots \int_{0}^{t_2}
\frac{dt_{1}}{\xi}\right]
 \left[\int_{0}^{t}\frac{dt_{n_-+1}}{\xi} \int_{0}^{t_{n_-+1}} \frac{dt_{n_-+2}}{\xi} ....
 \int_{0}^{t_{n_-+n_+ -1}} \frac{dt_{n_++n_-}}{\xi} \right]
\nonumber \\
& \left( \Delta_T \xi \right)^{n_-+n_+} \exp \left[ I_0(\{t_k\})\right]\nonumber
\end{eqnarray}
The interaction $I_0$ is obtained by evaluating the $Tr_{el}$, in other words
by solving the Keldysh problem of electrons in the time dependent potential
specified by the spin flips.  In principle this is a multiparticle interaction depending
on all of the  times $\{t_k\}$.  In the equilibrium problem an essentially
complete analytical solution exists \cite{Anderson69}, showing that  the interaction is pairwise,
with a logarithmic time dependence and coefficients given by the product of
the  sign of the charges (i.e. whether
the spin flips from up to down) and the  changes
in scattering phase shifts.    In the nonequilibrium case
general analytical expressions are not known. The available evidence, including
solutions at times short and long compared to $\Delta \mu$,
perturbative calculations \cite{mitra06} and numerics \cite{segal06}
suggests the following structure, which is slightly more involved than
in equilibrium. 

To specify the structure, it is convenient to collapse the two-contour
problem onto a single time axis by defining
classical ($V_{cl}= \frac{\lambda D}{2} \{S_z(t_-) + S_z(t_+)\}$) and quantum fields 
($V_q =\frac{\lambda D}{2} \{S_z(t_-) - S_z(t_+)\}$). We then have a four state system,
in which either $V_q=0$ and $V_{cl}=\pm \lambda D/2$ or $V_{cl}=0$ and $V_q=\pm \lambda D/2$.
Transitions between these states are instantons, which we 
label by an integer $n=\pm 1$ giving the sign of the change in the
quantum field (this is just the usual Coulomb gas charge)
and an integer $q=\pm1$ corresponding to the sign of the quantum field
in the region where it is non-zero. The right hand panels of Fig \ref{kdq} show some examples.

There are in principle 16 pairwise interactions between the 4 kinds of instantons, but
calculations reveal a simpler structure (see Appendix $A$). 
We find that the interaction between instantons at times $t_i$ and $t_j$  is pairwise,
as in equilibrium. As in equilibrium the
sign of the interaction between instantons $i$ and $j$ is
determined in the usual way by the product of the charges $n_in_j$, 
while the  magnitude is  determined by the  product of the 
scattering phase shifts. However in the general nonequilibrium case
the phase shifts are complex
and take value $\delta_i$ if $q_i=1$ and $\delta^*_i$ if $q_i=-1$ (see Appendix $A$). Finally,
with each interaction is a phase factor. The interaction has a logarithmic time dependence, given approximately
by $ln\frac{ia_{ij}|t_i-t_j|+\xi}{\xi}$, with phase factor $a_{ij}=\pm 1$ determined
by the {\it Keldysh} time-ordering of $t_i$ and $t_j$, so that in the upper
panel of  Fig~\ref{kdq} $a_{23}=1$ whereas in the middle panel $a_{23}=-1$. We find
it convenient to combine the phase factors into an over-all phase $\Phi$. The result is 
 
 \begin{eqnarray}
\langle \sigma |\hat{\rho}_S(t) |\sigma\rangle &= \sum_{n_- + n_+= even} 
(-i)^{n_-} i^{n_+}
\langle \sigma | \tau_x^{n_-}\rho_{S0} \tau_x^{n_+} |\sigma \rangle 
 \int_{0}^{t}\frac{dt_{n_-+n_+}}{\xi} \int_{0}^{t_{n_-+n_+ -1}}
 \ldots \int_{0}^{t_2} dt_1  \left( \Delta_T \xi \right)^{n_-+n_+} \label{couleq2}\\ 
& \sum^{\prime}_{n_i, q_i v_i}
\exp \left[\sum_{j<i} n_i n_j  C_0(q_i,q_j,\frac{t_i - t_j}{\xi}, 
\Delta \mu \xi) + i \Phi\right]
\nonumber
\end{eqnarray}

The prime symbol on the sum above is to keep track of
constraints such as, two charges of the same
sign cannot appear more than twice in sequence, and 
$\sum_i n_i = 0$ for an expansion involving the diagonal component of the density matrix. 

The long \cite{Ng96}  and short \cite{Anderson69,Nozieres69}
time limits of the interaction function $C_0$ are known.
At short times $t<1/\Delta \mu$
\begin{equation}
C_0(\Delta \mu t \ll 1) = N\left(\frac{\delta_L + \delta_R}{\pi}\right)^2\ln 
\left(\frac{|t|}{\xi}\right) \label{vtless}
\end{equation}
independent of quantum fields. 
At long times 
$t > \frac{1}{\Delta \mu}$,  we have
\begin{eqnarray}
C_0(q=+,q=+,\Delta\mu t>1;\Delta \mu \xi \ll 1)(t) = \{C_0(q=-,q=-)\}^*
&=&N\left(\frac{\delta_L^2}{\pi^2}+\frac{\delta_R^2}{\pi^2}\right)
\ln{\left(|\Delta\mu t|\right)}
+\Gamma_{neq}|t| \nonumber \\
 &+& N\left( \frac{\delta_{L} + \delta_R}{\pi}\right)^2
\left[ \ln{\frac{1}{\Delta\mu\xi}}\right]\label{vtgreat} \\
C_0(q=+,q=-,\Delta\mu t>1;\Delta \mu \xi \ll 1)(t) = C_0(q=-,q=+)
&=&N\left(\frac{|\delta_L|^2}{\pi^2}+\frac{|\delta_R|^2}{\pi^2}\right)
\ln{\left(|\Delta\mu t|\right)}
+\Gamma_{neq}|t| \nonumber \\
 &+& N\left( \frac{\delta_{L} + \delta_R}{\pi}\right)^2
\left[ \ln{\frac{1}{\Delta\mu\xi}}\right]\label{vtgreat1}
\end{eqnarray}
From Eqns.~\ref{vtgreat} and ~\ref{vtgreat1}, note that the dynamics is characterized
by an exponential time decay, which we use to define the {\em
decoherence rate} $\Gamma_{neq}$ which plays a fundamental role in the subsequent analysis:
\begin{equation}
\Gamma_{neq} = \gamma_{neq} \Delta \mu  = N\Delta\mu
\frac{|\delta_L^{\prime \prime}- \delta_R^{\prime
\prime}|}{2\pi}
= N \Delta \mu \xi \left(\frac{1}{\pi}\lambda^2\cos^2\theta_L \sin^2\theta_L +\ldots \right).
\label{gammaneq}
\end{equation}
.

The physics expressed by the interaction $C_0$ in the $\Delta \mu
\xi \ll 1$ limit is as follows: for times less than the dephasing
scale $t_{dephasing} = (\Delta \mu)^{-1}$ one has the equilibrium
result: the two level system interacts with one coherent
combinations of the two leads ($\bar{c}^{\dagger}= \cos\theta_L
c_L^{\dagger} + \sin\theta_R c_R^{\dagger}$); the other combination
decouples. The coupling leads to the usual power law interaction
with exponents given by the coherent phase shift $\delta_{eq} =
\delta_L + \delta_R$. Note that $\delta_{eq}$ is independent
of quantum fields.  For times longer than the dephasing scale, one
has a richer structure. The model is effectively a
two channel model with separate couplings to left
and right leads.  The interaction between instantons acquires
a dependence on the quantum field. At times longer than the decoherence scale
$\Gamma_{neq}^{-1}$ the interaction is cut off altogether.

These limiting
forms suggest the following decomposition of the interaction:

\begin{eqnarray}
C_0(q,q^{\prime},t-t^{\prime},\Delta \mu,\xi)= Q_0(q,q^{\prime}) h_0\left(\frac{t-t^{\prime}}{\xi}\right) + 
Q_M(q,q^{\prime})
h_M\left(\frac{t-t^{\prime}}{\xi},\Delta \mu (t-t^{\prime})\right) +
\gamma_{neq}h_{neq}\left(\frac{t-t^{\prime}}{\xi}, \Delta \mu (t-t^{\prime}) \right)
\label{C}
\end{eqnarray}
where the ``charges'' or phase shifts $Q_{0,M}$ obey the property
\begin{eqnarray}
Q_{0,M}(q,q^{\prime}) = \left[Q_{0,M}(-q,-q^{\prime}) \right]^* \label{quantsym1}\\
Q_{0,M}(q,-q) = Q_{0,M}(-q,q) => Im[Q_{0,M}(q,-q)] = 0 \label{quantsym2}
\end{eqnarray}
The precise form of the functions $h_{0,M,neq}$ depend on the cutoff
scheme, but for $t/\xi \gg 1 $ and $\Delta \mu \xi \ll 1 $ (see Appendix ~\ref{hfunc})
\begin{eqnarray}
h_0(t/\xi) &=& \ln(|t|/\xi)  \\
h_M(t/\xi,\Delta \mu t) &=& h_0(t/\xi), \,\,\,\,\,\, \Delta \mu t \ll 1 \\
&=& \ln \frac{1}{\Delta \mu \xi}, \,\,\,\,\,\, \Delta \mu t \gg 1\\
h_{neq}(t/\xi, \Delta \mu t) &=& \frac{2}{\pi} (\Delta \mu t)^2, \,
\,\,\,\,\, \Delta \mu t \ll 1 \\
&=& |\Delta \mu t|,  \,\,\,\,\,\, \Delta \mu t \gg 1
\end{eqnarray}
implying that for a model with $\Delta \mu \xi \ll 1$,
\begin{eqnarray}
Q_0(+,+)=N(\frac{\delta_L^2}{\pi^2}+\frac{\delta_R^2}{\pi^2}) \label{Q0++}\\
Q_M(+,+)= \frac{2N}{\pi^2}
\delta_L \delta_R \label{QM++}\\
Q_0(+,-) = N(\frac{\delta_L \delta_L^*}{\pi^2} 
+ \frac{\delta_R \delta_R^*}{\pi^2}) \label{Q0+-} 
\\ 
Q_M(+,-) = N \frac{\delta_L \delta_R^* + \delta_R \delta_L^*}{\pi^2}
\label{QM+-}
\end{eqnarray} 
The coefficient $\gamma_{neq}$ is independent of the
quantum fields and is given by 
\begin{equation}
\gamma_{neq}= N \frac{|\delta_L^{\prime \prime}-
\delta_R^{\prime \prime}|}{2 \pi}
\end{equation}

The first term proportional to $Q_0$ in Eq~\ref{C} represents the
effect of processes in which an electron emerges from one lead and
is scattered back into the same lead; it is independent of $\Delta
\mu$. The term proportional to $Q_M$ represents the effect of processes
in which an electron is transferred from one lead to another; it
depends on $\Delta \mu$. Finally, as will be seen, the last term
expresses the decoherence; it vanishes if the coupling is only to
one lead.  

It is also interesting to consider the expression for $C_0$ in
a model in which $\Delta \mu \xi > 1$. This situation arises after
rescaling. In this case the regime $\Delta \mu t < 1$ is not
defined, and for all times $t > \xi^{-1}$ we find,
\begin{eqnarray}
C_0(q=+,q=+,\Delta\mu t>1)
=Q_0
\left[\ln{\frac{ |t|}{\xi}} \right]
\label{dlessv}
\end{eqnarray}
with (for the model with $\Delta \mu \xi > 1$) $Q_M=0$ and
\begin{equation}
Q_0 \rightarrow Q_0^{\prime} = \left(\frac{\tan^{-1}{\frac{\lambda \cos^2\theta_L}{1-\frac{\lambda^2}{4}} }}{\pi} \right)^2 + \left( \frac{\tan^{-1}{\frac{\lambda \cos^2\theta_R}{1-\frac{\lambda^2}{4}} }}{\pi}\right)^2 \label{qoless}
\end{equation}
In this limit there is no decay because the theory has no real
process which allows nonconservation of energy (see Fig.~\ref{VD},
case $C$).

Comparison of Eqs \ref{vtgreat} and \ref{dlessv}  reveals an
important point. If we apply the usual renormalization process of
reducing bandwidth we must pass from the model which gives rise to
Eq \ref{vtgreat} and contains decoherence to the model which gives
rise to Eq \ref{dlessv}. No decoherence processes exist in this
latter model, but the renormalization maps one model onto another
model with the same physical content. We therefore conclude that one
consequence of renormalization must be the generation of a decay
rate, which appears as an extra parameter, additional to what is
directly
computed from the small bandwidth model.

\section{Derivation of scaling equations} \label{Scaling}

\begin{figure}[t]
\begin{center}
\includegraphics
[width=0.5\columnwidth]{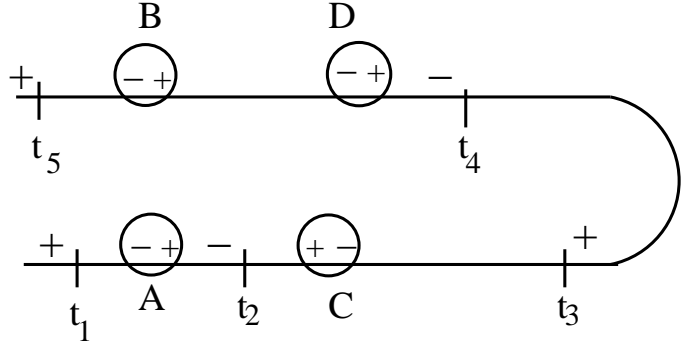} 
\end{center}
\caption{Scheme for integrating out close pairs.}
\label{cpinst}
\end{figure}

In the formulation given in Eq.~\ref{couleq2}, the nonequilibrium
two level system is seen to be a function of the dimensionless
parameters $\Delta_T \xi$, $\Delta \mu \xi$ and $\delta_{L,R}$. In
this section we construct a renormalization group analysis  by
following the usual procedure of reducing the energy cutoff, i.e.
increasing the time cutoff from  $\xi$ to $\xi'=(1+\Lambda) \xi$,
integrating out the degrees of freedom in the eliminated interval
and determining the consequences for the remaining degrees of
freedom. These effects were considered for the equilibrium problem
by \cite{Anderson69}: reducing the energy cutoff (increasing the
time cutoff) leads to changes arising from the $\xi$ dependence of
$C_0$. Rescaling leads to the simple ``engineering dimension" changes
$(\Delta_T,\Delta \mu)\rightarrow (\Delta_T,\Delta \mu)(\xi/\xi')\approx
(\Delta_T,\Delta \mu)(1-\Lambda) $. Finally, some kink-antikink pairs
fall within the time interval between $\xi$ and $\xi(1+\Lambda)$ and
must therefore be removed from the renormalized theory.  The
procedure for the nonequilibrium problem is  similar to the
equilibrium problem,
in that in the
starting formulation the minimum separation in time between
tunneling events 
is $\xi$. In the rescaled
theory, spin flip events separated by time intervals less than
$\xi'$ cannot explicitly appear, but their presence will lead to a
renormalization of the interaction between the processes which do
explicitly appear in the theory. For small $\Delta_T \xi \Lambda$ the
sequence of close tunneling events which appears with any
probability are the ``close pairs" shown in Fig.~\ref{cpinst}. A close pair may
lie between two spin-flip events that occur on the same Keldysh axis (example
pairs $A$  and $B$ in Fig~\ref{cpinst} that lie between spin-flip events $t_1,t_2$), 
or may lie between spin-flip events on opposite
Keldysh axes (examples $C$ and $D$ that lie between spin-flip events $t_2,t_4$).   
Physically, a close pair corresponds to a dipole, and leads to a
screening of the interaction between other spin flip events that may lie
on the same Keldysh axis or on different Keldysh axes.  Note that we always 
consider close pairs that lie on the same axis. A close pair 
with one member on each contour  cannot be considered as a dipole;
its removal would change the spin state at the final time $t$ or initial time $t=0$.

\begin{figure}[t]
\begin{center}
\includegraphics
[width=0.5\columnwidth]{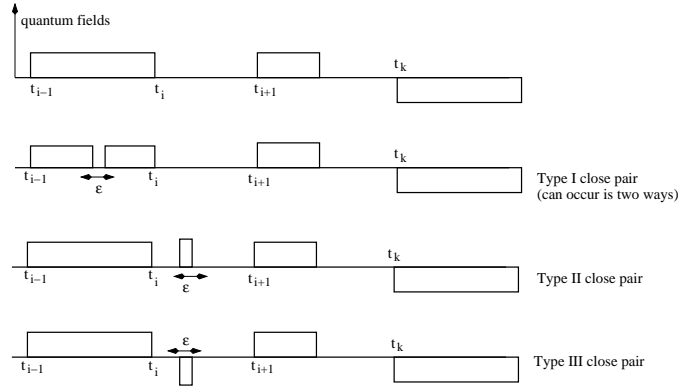} 
\end{center}
\caption{Close pairs that are nearest neighbors to charge at time $t_i$}
\label{rgcp}
\end{figure}

Mathematically, the calculation is easiest to perform in the collapsed single axis representation 
in terms of classical and quantum fields. As an example consider Fig.~\ref{rgcp}.
The interaction between the charge at $t_i$ and the charge at $t_k$ can get renormalized
by integrating out eight kinds of close pairs, four of which are nearest neighbors 
to charge at $t_i$, while the other four are nearest neighbors to the charge at
$t_k$. We show the four close pairs that are nearest neighbors to $t_i$ in Fig~\ref{rgcp}. 
Close pair I can occur in two ways which have the same quantum field configuration,
but different classical field configuration. 

The change in the density matrix, on integrating out the four close pairs is (${\bar \Delta}_T=i\Delta_T \xi$)
\begin{eqnarray}
\rho^{2n+2} = \rho^{2n} \left[1 + \bar{\Delta}_T^2\left(2*I + II + III \right) \right]
\label{rhocorr}
\end{eqnarray}
Summing over all possible positions of the relative separation $\epsilon$ 
between the two charges of the close pair, and their center
of mass position $\tau$ one obtains,
\begin{eqnarray}
I = \int_{t_{i-1}}^{t_i} 
\frac{d\tau}{\xi}\int_{\xi}^{\xi(1+\Lambda)}\frac{d\epsilon}{\xi}\exp{\left[- C_0(+,+,\epsilon)
+ \sum_{k}n_i n_k\left(-C_0(+,q_k, |t_k - \tau -\frac{\epsilon}{2}|)
+C_0(+,q_k|t_k - \tau +\frac{\epsilon}{2}|) \right)\right]} \\
II = \int_{t_{i}}^{t_{i+1}} 
\frac{d\tau}{\xi}\int_{\xi}^{\xi(1+\Lambda)}\frac{d\epsilon}{\xi}\exp{\left[-C_0(+,+,\epsilon)
+\sum_{k}n_i n_k\left(C_0(+,q_k, |t_k - \tau -\frac{\epsilon}{2}|)
-C_0(+,q_k|t_k - \tau +\frac{\epsilon}{2}|) \right)\right]} \\
III = \int_{t_i}^{t_{i+1}} 
\frac{d\tau}{\xi}\int_{\xi}^{\xi(1+\Lambda)}\frac{d\epsilon}{\xi}\exp{\left[- C_0(-,-,\epsilon)
+ \sum_{k}n_i n_k\left(-C_0(-,q_k, |t_k - \tau -\frac{\epsilon}{2}|)
+C_0(-,q_k|t_k - \tau +\frac{\epsilon}{2}|) \right)\right]}
\end{eqnarray}
In the above three expressions, the first term in the argument of the exponent represents 
the self interaction of the close pair, and
the second term represents the interaction of the close pair with all other charges. The latter may be
Taylor expanded in $\epsilon$. The term I on Taylor expansion takes the form
\begin{eqnarray}
I \sim 
\int_{t_{i-1}}^{t_{i}} \frac{d\tau}{\xi}
\int_{\xi}^{\xi(1+\Lambda)}\frac{d\epsilon}{\xi}e^{-C_0(+,+\epsilon)}
\left(1 - \epsilon \sum_{k}n_i n_k \frac{\partial
C_0(|t_k - \tau|)}{\partial \tau} \right) \label{cint}
\end{eqnarray}
The integrals over $\tau$ and $\epsilon$ can easily be performed. The result is 
of $O(\Lambda)$, which implies that the change to the density matrix in Eq.~\ref{rhocorr} is 
$O(\bar{\Delta}_T^2 \Lambda)\ll 1 $, and therefore can be re-exponentiated. Finally, integrating out
the four close pairs of Fig~\ref{rgcp} leads to the following renormalization of the density matrix, 
\begin{eqnarray}
&\rho^{2n+2} = \rho^{2n} e^{\bar{\Delta}_T^2 e^{-C_0(\xi)}\Lambda (\frac{- 2t_{i-1} + 2t_{i+1}}{\xi})}\\
&e^{\Lambda \bar{\Delta}_T^2 e^{-C_0(\xi)}
\sum_{k}n_i n_k 
\left[-2 C_0(+,q_k,|t_k - t_i|) + 2 C_0(+,q_k, |t_k - t_{i-1}|) 
+C_0(+,q_k, |t_k - t_{i+1}|) - C_0(+,q_k, |t_k - t_i|) - C_0(-,q_k,|t_{i+1}-t_k|) 
+ C_0(-,q_k,|t_k-t_i|) \right]}
\nonumber 
\end{eqnarray}
Note that the term $e^{-C_0(\xi)}$ is from the self-interaction of the close pair. Since
this term arises at short times $t\sim \xi$, it has no explicit quantum field dependence, and therefore
the quantum field label has been dropped.
 
Repeating the above computation for all possible positions of close pairs, 
one finds that the initial factor of
$e^{\bar{\Delta}_T^2 e^{-C_0(\xi)}\Lambda
(\frac{2t_{i+1}-2t_{i-1}}{\xi})}$ cancel among each other, while the
function $C_0(q_i,q_k,|t_k - t_i|)$
is renormalized in the following way
\begin{eqnarray}
&C_0(q_i,q_k,|t_k - t_i|) \rightarrow C_0(q_i,q_k,|t_k - t_i|) \label{dd}\\
&- \Lambda \bar{\Delta}_T^2
\left[6 C_0(q_i,q_k,|t_k-t_i|)
-C_0(-q_i,q_k,|t_k-t_i|) -C_0(q_i,-q_k,|t_k-t_i|) \right] e^{-C_0(\frac{t}{\xi}=1)} \nonumber
\end{eqnarray}
Note that $C_0(t/\xi=1)=0$ if $\Delta \mu \xi <<1$ but
$=\Gamma_{neq}\xi$ if $\Delta \mu \xi >1$.

\begin{figure}[t]
\begin{center}
\includegraphics[width=0.5\columnwidth]{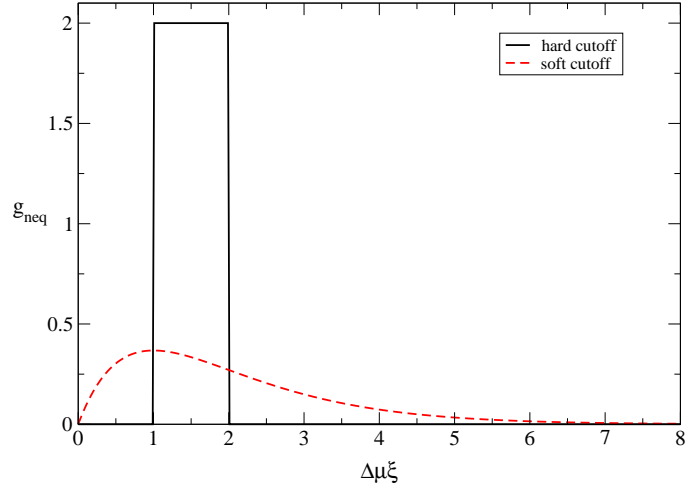}
\caption{Crossover function $g_{neq}$ (in units of $\frac{\pi}{2}Q_M \Delta \mu$) 
describing additive renormalization of decay constant
$\Gamma_{neq}$, for hard 
(solid line) and soft (dashed line)
cutoffs.} \label{crossover}
\end{center}
\end{figure}

The second effect arises from the explicit dependence of the interaction
$C_0$ on the cutoff $\xi$. We write, for infinitesimal $\Lambda$,
\begin{equation}
C_0(t,\xi,\Delta\mu)= C_0(\frac{t}{\xi^{\prime}},\Lambda \Delta\mu
t)+\delta C_0 \label{dc}
\end{equation}
with
\begin{equation}
\delta C_0=\frac{dC_0(\frac{t}{\xi},\Delta\mu t)}{d ln \xi}
\label{delC}
\end{equation}
Adding Eqns~\ref{dd} and \ref{dc} leads to the fundamental scaling
equation,
\begin{equation}
\frac{dC_0(q,q^{\prime},t)}{d\ln \xi} = -\bar{\Delta}_T^2 e^{-C_0(1)}\left[6 C_0(q,q^{\prime},t)
-C_0(-q,q^{\prime},t) -C_0(q,-q^{\prime},t) \right]
+\frac{d C_0(q,q^{\prime},t/\xi,\Delta\mu \xi)}{d \ln \xi} \label{scmain}
\end{equation}
The second term in Eq.~\ref{scmain} may be computed from the
fundamental  Eq.~\ref{closedloop}. While a precise general expression
is not known, the limits are established. For $\Delta \mu \xi \ll 1$,
Eqns.~\ref{vtgreat} and~\ref{vtgreat1} show that $\frac{d C_0}{d \ln \xi}$
is independent of time, with coefficient $Q_0 + Q_M$ independent of quantum fields. This
is just the equilibrium scaling. For $\Delta \mu \xi \gg 1 $, Eq.~\ref{dlessv} shows that 
$\frac{d C_0}{d \ln \xi}$ is again independent of time, with some coefficient $Q_0$ 
derived from scaling and not, in general, given by Eq.~\ref{qoless}. The regime $\Delta \mu \xi \sim 1$
requires a more careful treatment. Differentiation of the perturbative results derived in 
Appendix ~\ref{hfunc} yields two contributions: a function $g_M = \frac{d h_M}{d \ln \xi}$
which expresses the dephasing crossover by "turning off" 
the $Q_M$ contribution to $C_0$ as $\Delta \mu \xi$ increases through
unity, and an additional contribution proportional to $t$ for $\Delta \mu t > 1$
which expresses decoherence by cutting off the interaction between instantons.
The decoherence term is characterized by a  
coefficient 
$g_{neq} = Lt_{t\rightarrow \infty} \left[ \gamma_{neq} \frac{1}{t}\frac{dh_{neq}}{d\ln\xi}\right]$.
Fig~\ref{crossover} shows 
$g_{neq}$ calculated from Appendix ~\ref{hfunc} within perturbation theory to ${\cal O}(\lambda^2)$ and 
for hard and soft cutoffs. (Note that the hard cutoff model gives rise to oscillations
which we have neglected).
 
In summary, our scaling theory must keep track of the changes in the function
$C_0$ as the cutoff is changed, but consideration of the short and
long time limits shows that the model has effectively six coupling
constants: ``Coulomb gas charges" $Q_0(+,+), Q_0(+,-)$ expressing the part of logarithmic
interaction between tunneling events involving the same lead, Coulomb gas charges
$Q_M(+,+),Q_M(+,-)$ expressing the part of the
logarithmic interaction which depends on coherence between leads, a
tunneling amplitude $\Delta_{Tq}$ (which acquire a quantum field
dependence labeled by $q$), and the decoherence rate $\Gamma_{neq}$.
These are renormalized according to
\begin{eqnarray}
\frac{d\bar{\Delta}_{T\pm}}{d\ln\xi} &=&\bar{\Delta}_{T\pm}\left[1-\frac{1}{2}\left[Q_0(\pm,\pm)+
Q_M(\pm,\pm) g_{M}(\Delta \mu \xi)\right]\right] \label{scq1}\\
\frac{d\Gamma_{neq}(+,+)}{d ln \xi} &=&-4\left[\Gamma_{neq}(+,+)\bar{\Delta}_{T+}^2+ \frac{1}{2}\left(\Gamma_{neq}(+,+)
 \bar{\Delta}_{T+}^2 -\Gamma_{neq}(+,-) \bar{\Delta}_{T-}^2 \right)
\right] e^{-C_0(1)} + g_{neq}(\Delta\mu \xi) \label{scq2}\\
\frac{d\Gamma_{neq}(-,-)}{d ln \xi} &=&-4\left[\Gamma_{neq}(-,-)\bar{\Delta}_{T-}^2+ \frac{1}{2}\left(\Gamma_{neq}(-,-)
 \bar{\Delta}_{T-}^2 -\Gamma_{neq}(-,+) \bar{\Delta}_{T+}^2 \right)
\right] e^{-C_0(1)} + g_{neq} 
(\Delta\mu \xi) \label{scq3}\\
\frac{d\Gamma_{neq}(+,-)}{d ln \xi} &=&-\left[3\Gamma_{neq}(+,-)\bar{\Delta}_{T-}^2 + 3\Gamma_{neq}(-,+)
\bar{\Delta}_{T+}^2 -\Gamma_{neq}(+,+) \bar{\Delta}_{T+}^2-
\Gamma_{neq}(-,-) \bar{\Delta}_{T-}^2 \right] e^{-C_0(1)} \nonumber \\
&+& g_{neq}(\Delta\mu \xi)\label{scq4}\\
\frac{dQ_{0,M}(+,+)}{d ln \xi} &=&-4\left[Q_{0,M}(+,+)\bar{\Delta}_{T+}^2+ \frac{1}{2}\left(Q_{0,M}(+,+)
 \bar{\Delta}_{T+}^2 -Q_{0,M}(+,-) \bar{\Delta}_{T-}^2 \right)
\right] e^{-C_0(1)} \label{scq5} \\
\frac{dQ_{0,M}(+,-)}{d ln \xi} &=&-\left[3Q_{0,M}(+,-)\bar{\Delta}_{T-}^2 + 3Q_{0,M}(-,+)\bar{\Delta}_{T+}^2 
-Q_{0,M}(+,+) \bar{\Delta}_{T+}^2-
Q_{0,M}(-,-) \bar{\Delta}_{T-}^2 \right] e^{-C_0(1)} \label{scq6}\\
\frac{dQ_{0,M}(-,-)}{d ln \xi} &=&-4\left[Q_{0,M}(-,-)\bar{\Delta}_{T-}^2+ \frac{1}{2}\left(Q_{0,M}(-,-)
 \bar{\Delta}_{T-}^2 -Q_{0,M}(-,+) \bar{\Delta}_{T+}^2 \right)
\right] e^{-C_0(1)} \label{scq7}
\end{eqnarray}

Note that in these expressions $C_0(1)=\Gamma_{neq}\xi$ for $\Delta
\mu \xi \gg 1$ and $=0$ otherwise. 
The meaning of Eq.~\ref{scq2},~\ref{scq3},~\ref{scq4} is that as the renormalized
chemical potential passes through the scale $\Delta \mu \xi =1$ an
additive contribution to $\Gamma_{neq}$ is generated.

\section{Solution of the scaling equations} \label{Solution}

We now discuss the solution of the scaling equations.
Notice from Eq.~\ref{Q0++} -~\ref{QM+-} 
that in equilibrium ($\Delta \mu = 0$),
the charges $Q_{0,M}$ are independent of the quantum fields. Moreover
out of equilibrium, any explicit dependence on the quantum fields appears
to $O(\lambda^3)$ for asymmetric couplings, and to $O(\lambda^4)$ for symmetric
couplings. We will present results that are valid to $O(\lambda^2)$, and hence
ignore the explicit quantum field dependence of the charges. In this limit, as
we shall show, the dominant effect of voltage is due to the decoherence term. 

The physically relevant starting point for the solution of the scaling
equations is one where the chemical potential difference is small
compared to the cutoff scale. 
In this limit $C_0(1),g_{neq}\approx 0$,  and $g_{M}(\Delta \mu
\xi) \approx 1$; thus the effective Coulomb gas charge $Q=Q_0+Q_M$
(physically, the level couples to the coherent combination of the
leads). The scaling equations thus become
\begin{eqnarray}
\frac{d\bar{\Delta}_T}{d\ln\xi} &=&\bar{\Delta}_T\left[1-\frac{Q}{2}\right] \label{sceq2}\\
\frac{dQ}{d ln \xi} &=&-4Q\bar{\Delta}_T^2  \label{Q} \\
\frac{d\Gamma_{neq}}{d ln \xi} &=& -4\Gamma_{neq} \bar{\Delta}_T^2 \label{sceq3}\\
\frac{dQ_M}{d ln \xi}&=&-4Q_M\bar{\Delta}_T^2  \label{QMeq}
\end{eqnarray}
Eqs \ref{sceq2} and \ref{Q} are the usual equilibrium scaling
equations and are solved as usual; from the solution the behavior of
$Q_M$ and $\Gamma_{neq}$ is computed. The equations leave the
combination $I=\bar{\Delta}_T^2-\frac{1}{2}(\frac{Q-2}{2}-ln(Q/2))$ invariant. If
$I<0$ then the model scales toward $\Delta_T=0$ (i.e. is localized)
while if $I>0$ then it is on the delocalized side of the phase
boundary and $\Delta_T$ increases. We discuss the two cases
separately.

If the initial conditions are such that the model is on the
localized side of the equilibrium phase diagram, then scaling
proceeds until  the cutoff crosses through the dephasing scale
$\Delta \mu \xi =1$. Beyond this point, changes occur. First, the
leads decohere: $g_{M} \rightarrow 0$ so the term proportional
to $Q_M$ drops out of the scaling equations and the charge becomes
$Q_0$. Depending on the sign of $Q_M$ (i.e. the relative sign of
$\delta_R$ and $\delta_L$) this may either make the system more
localized or more delocalized. Second, and of greater significance,
the decoherence rate $\Gamma_{neq}$ acquires a positive
additive term, arising from the function $g_{neq}$ in the scaling
equation. 
Third, the effective coupling becomes
$(\bar{\Delta}_T^*)^2\equiv\bar{\Delta}_T^2e^{-C_0(1)}\approx
\bar{\Delta}_T^2e^{-\Gamma_{neq} \xi}$. Thus the theory at the scale
$\xi>(\Delta \mu)^{-1}$ is characterized by a fugacity $ln
\Delta_T^*\xi$, a decoherence rate $\Gamma_{neq}$, a charge $Q_0$,
and by the scaling equations
\begin{eqnarray}
\frac{d\bar{\Delta}_T}{d\ln\xi} &=&\bar{\Delta}_T\left[1-\frac{Q_0}{2}\right] \label{scmu1}\\
\frac{d\Gamma_{neq}}{d ln \xi} &=& -4\Gamma_{neq} \bar{\Delta}_T^2 e^{-C_0(1)}  \label{scmu2}\\
\frac{dQ_0}{d ln \xi} &=&-4Q_0\bar{\Delta}_T^2 e^{-C_0(1)} \label{Q0mu} 
\end{eqnarray}
We see that scaling proceeds in the usual way until the {\em
nonequilibrium scale} $\xi=1/\Gamma_{neq}$ is reached; beyond this
point the factor $ e^{-C_0(1)}$ cuts off the scaling and we are left
with a perturbative theory.

A particularly important special case occurs if $I=0$. In this case
the equilibrium fixed point is $\Delta_T=0$, $Q=2$, and if the initial
value of $\Delta \mu $ is sufficiently small, this fixed point is
approached very closely, so that at the scale $\xi=1/\Delta \mu$
\begin{eqnarray}
\bar{\Delta}_T(\xi)&=&\frac{\bar{\Delta}_{T0}}{1+2\bar{\Delta}_{T0}ln\frac{1}{\Delta
\mu \xi_0}}\approx \frac{1}{2 ln\frac{1}{\xi_0\Delta \mu}}
\label{Dcritical} \\
Q&\approx& 2 \\
Q_M &\rightarrow& Q_M^*
\end{eqnarray}

Scaling through the crossover region then drives $Q_M \rightarrow
Q_M^{*}$, changes the basic charge to $Q_0=2-Q_M^*$, generates a
$\Gamma_{neq}=\frac{\pi}{2} \left|Q_M^*\right|\Delta \mu$, and does not change
$\Delta_T$ significantly. Scaling then proceeds until $\xi \rightarrow
1/\Gamma_{neq} \sim 2/(\pi \left|Q_M^*\right|\Delta \mu$). We therefore see
that the dephasing crossover typically shifts the system away from
the critical point, and that decoherence then cuts off the scaling.
The decoherence cutoff occurs very rapidly, unless $Q_M^* << 1$,
meaning that one of the leads is much more weakly coupled than the
other one. In this case a significant nonequilibrium scaling regime
can exist.

If the initial condition is on the delocalized side of the
equilibrium phase diagram, then again we distinguish two cases,
according to whether or not the model flows to strong coupling
before $\Delta \mu \xi =1$ or not. In the latter case, the treatment
outlined above applies. In the former case, the Kondo or coherence
scale is larger than the dissipation rate and a different treatment,
beyond the scope of this paper, is needed.

\section{Conclusions \label{Conclusion}}

In this paper we have expressed the nonequilibrium tunneling center model in terms of a
Coulomb gas defined on the Keldysh contour. The nonequilibrium problem
has a richer structure than the corresponding equilibrium problem. In particular
the effective low energy theory is shown to be a Coulomb gas
characterized by two parameters (charge and sign of quantum field). 
Crucial ingredients of the resulting theory are the dephasing arising because the wave
functions in the two leads precess at rates which differ by the
chemical potential difference, and a decoherence arising from the
dissipative processes again allowed when the model is driven out of
equilibrium. We showed explicitly that the decoherence effects cut off the power
law interaction between instantons which is found at $T=0$ in equilibrium.

Further, we generalized the standard equilibrium scaling theory of
the model to the nonequilibrium case. We found that scaling through
the dephasing crossover $\Delta \mu \xi \sim 1$ generates an 
additive renormalization to the decoherence rate. From this we
conclude that the decoherence rate is a fundamental parameter of the
nonequilibrium theory, which must be explicitly considered in a
renormalization process. We further showed explicitly how the
decoherence cuts off the renormalization group flow.

A few words on the generalization of this approach to other
quantum impurity models, such as the nonequilibrium Kondo model 
\cite{Paaske04a}. The key difference between
the model studied here and the nonequilibrium Kondo model is 
the term $\Delta_T S_x$ in Eq.~\ref{hloc} responsible for spin-flip processes.
The analog of $\Delta_T S_x$ for the Kondo model is $(J_{LL}+ J_{RR})(S_{+}s_{-} + h.c.)$
where $S$ is the impurity spin, while $\vec{s}= \sum_{\sigma \sigma^{\prime}k k^{\prime}}
\psi^{\dagger}_{k\sigma}\vec{\tau}\psi_{k^{\prime}\sigma^{\prime}} $ are the electron spins which have
been written as the following linear combination of the two leads $\psi_{k \sigma} = \frac{\sqrt{J_{LL}}}{\sqrt{J_{LL}+ J_{RR}}} c_{k
\sigma L} + \frac{\sqrt{J_{RR}}}{\sqrt{J_{LL}+ J_{RR}}} c_{k \sigma
R}$.
Thus in the
Anderson-Yuval-Hamann procedure applied to the Kondo model, the $n_{-}+ n_{+}$-th order expansion in 
the spin flip amplitude involves the computation of 
\begin{eqnarray}
&Tr_{electron}\left[ T e^{-i \int _0^t
dt^{\prime}\tilde{V}_{\uparrow}(t^{\prime})}s_-(t_{n_-})\ldots
s_+(t_1) \rho_{S0}^{\uparrow} \tilde{T} e^{+i \int _0^t
dt^{\prime}\tilde{V}_{\uparrow}(t^{\prime})}s_-(t_{n_-+n_+})\ldots 
s_+(t_{n_-+1})\right] \nonumber \\
&= e^{C_0(t_1,t_2 \ldots t_{n_-+n_+})} L(t_1,t_2 \ldots t_{n_-+n_+})
\label{kondo1}
\end{eqnarray}
rather than the quantity  needed for the model studied in this paper
\begin{equation}
Tr_{electron}\left[ T e^{-i \int _0^t
dt^{\prime}\tilde{V}_{\uparrow}}\rho_{S0}^{\uparrow} \tilde{T} e^{+i \int _0^t
dt^{\prime}\tilde{V}_{\uparrow}}\right] =  e^{C_0(t_1,t_2 \ldots t_{n_-+n_+})}
\label{sb1}
\end{equation}
In equilibrium, the quantity $L$ in Eq.~\ref{kondo1}
(referred to in the literature as the open
line contribution) \cite{Nozieres69} acquires the same structure as that of the closed 
loop part $e^{C_0}$,
namely that of a Cauchy determinant. Thus 
Eqns.~\ref{kondo1} and Eq.~\ref{sb1} 
and therefore the Kondo model
and the tunneling center model may be related by a simple redefinition
of the phase shifts.
Out of equilibrium this analysis breaks down because the dephasing between the
leads occuring at $\Delta \mu t > 1$ gives rise to
a two channel 
structure similar
to that analyzed by Fabrizio et al.~\cite{gogolin} and Vlad\'{a}r et al.~\cite{vlada}. A direct 
numerical evaluation or a mapping to an explicit two channel Kondo
model (with decoherence) could
be employed.  

This work suggests several generalizations. First, the long-time
exponential cutoff suggests that a numerical estimation of the
perturbation series may be possible. Second, the key issue in seeing
a wide nonequilibrium scaling range is to get the decoherence time
to be very large compared to the dephasing time. This does not occur
naturally in the simple two lead models we have studied. A search
for models, involving for example three leads, where this separation
of scales occurs more naturally, may be  of interest. 

Wide classes of models
have been studied in equilibrium by Hubbard-Stratonovich methods, in which the 
partition function is expressed as a sum over configurations of auxiliary fields. 
In the strong coupling limit of many quantum impurity models a small number of auxiliary field
configurations are relevant and the physics is controlled by tunneling between them.~\cite{Hamann}
Generalizing this analysis to the nonequilibrium situation is an important open
problem~\cite{mitra06}, for which the methods developed here may be useful. A useful
first step might be a comparison to the Bethe-ansatz solvable interacting resonant level
model~\cite{Andrei}.

{\it Acknowledgments} 
This work was supported by NSF-DMF 0431350.

\appendix

\section{Derivation of Eq.~\ref{couleq}} \label{derivecg}

We start from the Hamiltonian in Eq.~\ref{hdef} which we write
as a sum of two parts:
\begin{eqnarray}
H &=& H_0 + H_1\\
H_0 &=& S_z B  + \lambda D S_z \sum_{\alpha =
1..N}d^{\dagger}_{\alpha } d_{\alpha} + \sum_{a=L,R,\alpha=1..N}\int
d\epsilon \frac{\epsilon}{D} c_{\epsilon a\alpha}^{\dagger}
c_{\epsilon a \alpha} + \sqrt{\frac{1}{\pi}}  \sum_{a
= L,R,\alpha=1..N}\int_{\mu_a - 1/\xi}^{\mu_a+1/\xi} d\epsilon\left( \cos\theta_{a} c_{\epsilon a
\alpha}^{\dagger} d_{\alpha} + h.c.\right) \nonumber \\
H_1 &=& \Delta_T S_x
\end{eqnarray}
When $\Delta_T=0$, the Hamiltonian is exactly solvable and represents
a noninteracting resonant level hybridized with the leads. The
Anderson-Yuval-Hamann approach involves a perturbative
expansion in $\Delta_T$, treating $H_0$ exactly. This procedure
was originally carried out for the partition function; we apply it here to the 
time-dependent density matrix $\rho(t)$ determined from an initial condition
$\rho_0(t)$ via

\begin{equation}
\rho(t) = e^{-i H t} \rho_0 e^{i H t }
\end{equation}
The reduced density matrix for the impurity spin is defined as
\begin{equation}
\rho_S(t) = Tr_{el} \rho(t) \label{rho1}
\end{equation}
and is a $2\times2$ matrix whose diagonal elements give the
probability of the spin $S$ to be up or down, while the off diagonal
elements contain information about phase coherence. Rewriting
\begin{eqnarray}
e^{-i H t } = e^{-i H_0 t} T e^{-i \int_0^t dt V(t) } \\
V(t) = e^{i H_0 t} H_1 e^{-i H_0 t} = \Delta_T \left[e^{i H_0 t} S_x
e^{-i H_0 t}\right]=\Delta_T S_x(t)
\end{eqnarray}
a perturbative expansion in $V(t)$ of Eq.~\ref{rho1} may be carried
out, yielding
\begin{eqnarray}
\hat{\rho}_S(t) &=& \sum_{n_-,n_+} (-i)^{n_-} i^{n_+}
\left( \Delta_T \xi \right)^{n_-+n_+}  \\
&& \left[\int_{0}^{t}\frac{dt_{n_-}}{\xi} \int_{0}^{t_{n_-}}
\frac{dt_{n_--1}}{\xi} \ldots \int_{0}^{t_2}
\frac{dt_{1}}{\xi}\right]
 \left[\int_{0}^{t}\frac{dt_{n_-+1}}{\xi} \int_{0}^{t_{n_-+1}} \frac{dt_{n_-+2}}{\xi} ....
 \int_{0}^{t_{n_-+n_+ -1}} \frac{dt_{n_++n_-}}{\xi} \right]
\nonumber \\
&& Tr_{el}\left[S_x(t_{n_{-}})\ldots S_x(t_2)S_x(t_1) \rho_0
S_x(t_{n_- + n_+}) \ldots S_x(t_{n_{-}+1})\right]\nonumber
\end{eqnarray}
We assume the initial
density matrix
\begin{equation}
\rho_0 = \begin{pmatrix}\rho_{0S}^{\uparrow} \rho^{\uparrow}_{leads}
&& 0
\\ 0 & \rho_{0S}^{\downarrow} \rho^{\downarrow}_{leads}
\end{pmatrix}\label{rhoin}
\end{equation}
where $\uparrow,\downarrow$ represent the direction of the local
impurity spin, while $\rho^{\uparrow / \downarrow}_{leads}$
represents the steady state distribution of the electrons when the
local spin is oriented along $\uparrow/ \downarrow$. The effect of
the spin flip term would be to modify the diagonal components of
$\rho_S$ from its initial value, and also to introduce off diagonal
terms. The perturbative expansion for the diagonal component of $
\rho_S$ is (note $H_0^{\uparrow/\downarrow}$ appearing below implies $H_0$ corresponding to
$S_z=\uparrow/\downarrow$),
\begin{eqnarray}
\langle \uparrow | \rho_S | \uparrow \rangle &=&
\sum_{n_-,n_+=0,2,4\ldots} (-i)^{n_-} i^{n_+}
\left( \Delta_T \xi \right)^{n_-+n_+}  \\
&& \left[\int_{0}^{t}\frac{dt_{n_-}}{\xi} \int_{0}^{t_{n_-}}
\frac{dt_{n_--1}}{\xi} \ldots \int_{0}^{t_2}
\frac{dt_{1}}{\xi}\right]
 \left[\int_{0}^{t}\frac{dt_{n_-+1}}{\xi} \int_{0}^{t_{n_-+1}} \frac{dt_{n_-+2}}{\xi} ....
 \int_{0}^{t_{n_-+n_+ -1}} \frac{dt_{n_++n_-}}{\xi} \right]
\nonumber \\
&&  Tr_{el}\left[e^{+i H_0^{\downarrow}t_{n-}}
e^{-iH_0^{\uparrow}t_{n-}}\ldots e^{+i H_0^{\downarrow}t_{1}}
e^{-iH_0^{\uparrow}t_{1}} \rho_{S0}^{\uparrow} e^{+i
H_0^{\uparrow}t_{n_- + n_+}} e^{-iH_0^{\downarrow}t_{n_-+n_+}}\ldots
e^{+i H_0^{\uparrow}t_{n_-+1}} e^{-iH_0^{\downarrow}t_{n_-+1}}
\right]\nonumber \\
&&+ \sum_{n_-,n_+=1,3,5\ldots} (-i)^{n_-} i^{n_+} \left( \Delta_T \xi
\right)^{n_-+n_+}  \nonumber\\
&& \left[\int_{0}^{t}\frac{dt_{n_-}}{\xi} \int_{0}^{t_{n_-}}
\frac{dt_{n_--1}}{\xi} \ldots \int_{0}^{t_2}
\frac{dt_{1}}{\xi}\right]
 \left[\int_{0}^{t}\frac{dt_{n_-+1}}{\xi} \int_{0}^{t_{n_-+1}} \frac{dt_{n_-+2}}{\xi} ....
 \int_{0}^{t_{n_-+n_+ -1}} \frac{dt_{n_++n_-}}{\xi} \right]\nonumber \\
&& Tr_{el}\left[e^{+i H_0^{\uparrow}t_{n-}}
e^{-iH_0^{\downarrow}t_{n-}}\ldots e^{+i H_0^{\uparrow}t_{1}}
e^{-iH_0^{\downarrow}t_{1}} \rho_{S0}^{\downarrow} e^{+i
H_0^{\downarrow}t_{n_- + n_+}} e^{-iH_0^{\uparrow}t_{n_-+n_+}}\ldots
e^{+i H_0^{\downarrow}t_{n_-+1}}
e^{-iH_0^{\uparrow}t_{n_-+1}}\right]\nonumber
\end{eqnarray}

The above may be written as
\begin{eqnarray}
\langle \uparrow | \rho_S | \uparrow \rangle &=
\sum_{n_-,n_+=0,2,4\ldots} (-i)^{n_-} i^{n_+}
\left( \Delta_T \xi \right)^{n_-+n_+}  \label{rhoexp}\\
& \left[\int_{0}^{t}\frac{dt_{n_-}}{\xi} \int_{0}^{t_{n_-}}
\frac{dt_{n_--1}}{\xi} \ldots \int_{0}^{t_2}
\frac{dt_{1}}{\xi}\right]
 \left[\int_{0}^{t}\frac{dt_{n_-+1}}{\xi} \int_{0}^{t_{n_-+1}} \frac{dt_{n_-+2}}{\xi} ....
 \int_{0}^{t_{n_-+n_+ -1}} \frac{dt_{n_++n_-}}{\xi} \right]
\nonumber \\
&  Tr_{el}\left[ T e^{-i \int _0^t
dt^{\prime}\tilde{V}_{\uparrow}(t^{\prime})} \rho_{S0}^{\uparrow}
\tilde{T} e^{+i \int _0^t
dt^{\prime}\tilde{V}_{\uparrow}(t^{\prime})}\right]\nonumber \\
&+ \sum_{n_-,n_+=1,3,5\ldots} (-i)^{n_-} i^{n_+} \left( \Delta_T \xi
\right)^{n_-+n_+}  \nonumber\\
& \left[\int_{0}^{t}\frac{dt_{n_-}}{\xi} \int_{0}^{t_{n_-}}
\frac{dt_{n_--1}}{\xi} \ldots \int_{0}^{t_2}
\frac{dt_{1}}{\xi}\right]
 \left[\int_{0}^{t}\frac{dt_{n_-+1}}{\xi} \int_{0}^{t_{n_-+1}} \frac{dt_{n_-+2}}{\xi} ....
 \int_{0}^{t_{n_-+n_+ -1}} \frac{dt_{n_++n_-}}{\xi} \right]\nonumber \\
& Tr_{el}\left[ T e^{-i \int _0^t
dt^{\prime}\tilde{V}_{\downarrow}(t^{\prime})}
\rho_{S0}^{\downarrow} \tilde{T} e^{+i \int _0^t
dt^{\prime}\tilde{V}_{\downarrow}(t^{\prime})} \right]\nonumber
\end{eqnarray}
where
\begin{equation}
\tilde{V}_{\uparrow/\downarrow}(t) = \lambda D \left[S_z(t) +
\frac{(-/+)1}{2}\right]\sum_{\alpha= 1 \ldots
N}d^{\dagger}_{\alpha}(t) d_{\alpha}(t) \label{scatpot}
\end{equation}
$S_z(t)$ in \ref{scatpot} is the expectation value of the
operator $S_z$ and switches between $(+/-)\frac{1}{2}$ at times
$t_{1},t_2 \ldots t_{n_-+n_+}$. 

The first term in Eq.~\ref{rhoexp} ($n_{\pm}=0,2,4 \ldots$) represents ``out-scattering'', the
second term ($n_{\pm}=1,3,5 \ldots$) represents ``in-scattering''. To evaluate the $Tr_{el}$
in the out-scattering terms we write the lead states in the basis of scattering states 
appropriate to the static potential $(\lambda D/2)\sum_{\alpha= 1 \ldots
N}d^{\dagger}_{\alpha}(t) d_{\alpha}(t)$. The potential $\tilde{V}_{\uparrow}(t)$
then alternates between the values $-\lambda D\sum_{\alpha= 1 \ldots
N}d^{\dagger}_{\alpha}(t) d_{\alpha}(t) $ and $0$. Similarly to evaluate the 
$Tr_{el}$ in the in-scattering term we write the lead states in the basis
of scattering states appropriate to the static potential $(-\lambda D/2)\sum_{\alpha= 1 \ldots
N}d^{\dagger}_{\alpha}(t) d_{\alpha}(t)$ so $\tilde{V}_{\downarrow}$ alternates between 
$\lambda D\sum_{\alpha= 1 \ldots
N}d^{\dagger}_{\alpha}(t) d_{\alpha}(t) $ and $0$. 
The density of states of the two scattering problems is identical.

Since, the leads are noninteracting electrons, with nonequilibrium
imposed by $\mu_a \neq \mu_b$, the evaluation of the
$Tr_{el}$ for a given configuration of spin-flips reduces to
a problem of single particle quantum mechanics in a time dependent
potential. We briefly outline the solution based on the 
nonequilibrium linked cluster theorem \cite{mitra06} which implies
\begin{equation}
Tr_{el}\left[ T e^{-i \int _0^t
dt^{\prime}\tilde{V}_{\uparrow}(t^{\prime})} \rho_{S0}^{\uparrow}
\tilde{T} e^{+i \int _0^t
dt^{\prime}\tilde{V}_{\uparrow}(t^{\prime})}\right] = e^{-C_0(t)}
\end{equation}
where
\begin{equation}
C_0(t) = \sum_{\alpha}\int_0^{1}\frac{d g}{g} \int_0^t
dt_1 Tr \left[ {\phi}_{g,q}(t_1) \{ {G}^{K g}_{\alpha}(t_1,t_1)
+{G}^{Z g}_{\alpha}(t_1,t_1) \}\right]\label{closedloop}
\end{equation}
where for in-scattering,
${G}^{K,Z g}_{\alpha}(t_1,t_1)$ are Greens functions appropriate 
to the classical field 
$\phi_{cl}= \frac{\lambda D}{2}\left(S_z(t_+) + S_z(t_-)+1 \right)$, and the quantum field
$\phi_{g, q}= \frac{g \lambda D}{2}\left(S_z(t_-) -
S_z(t_+)\right)$. They obey the Dyson equation
\begin{equation}
\hat{G}^{g} = \hat{g} +
\hat{g} \left( \phi_{cl} 1 + \phi_{g,q} \tau_x \right) \hat{G}^{g}
\label{dysona}
\end{equation}
where $\hat{G}^{g} = \begin{pmatrix} G^R_{g} & G^K_{g}\\ G^Z_{g} & G^A_{g} 
\end{pmatrix}$,
$\hat{g} = \begin{pmatrix} g^R & g^K\\ 0 & g^A \end{pmatrix}$.
Here $g^{R,A,K}$ are the
standard retarded, advanced and Keldysh Greens functions of $H$, Eq \ref{hdef}, with
$S_z=-1/2$.
Note that $g^{R,A}$ are short ranged in time  and may be 
approximated as delta functions

It is convenient to recast Eq \ref{dysona} as
\begin{equation}
\hat{G}^{g} = \hat{{\bar g}} +
\hat{{\bar g}}  \phi_{g,q} \tau_x \hat{G}^{g}
\label{dysona1}
\end{equation}
with
$\hat{{\bar g}} =\left(1-\phi_{cl} \hat{g}\right)^{-1}\hat{g}$.  
Explicitly, we have 
\begin{eqnarray}
{\bar g}^{R,A}(t-t') &\approx& -\frac{(\phi_{cl}(t)-\lambda D/2) \pm i D }{(\phi_{cl}(t)-\lambda D/2)^2 + D^2}\delta(t-t')
\label{gbarR} \\
{\bar g}^K(t,t') &=& {\bar g}^R h-h{\bar g}^A=
-\frac{2D\left(cos^2\theta_Le^{-i\mu_L(t-t^\prime)}+cos^2\theta_Re^{-i\mu_R(t-t^\prime)}\right)}
{\left(iD+\lambda D/2-\phi_{cl}(t)\right)\left(-iD+\lambda D/2-\phi_{cl}(t')\right)}
P\frac{1}{t-t^\prime}
\label{bargk}
\end{eqnarray}
with $h\sim 1/(t-t')$ the usual distribution function. Note that for times $t,t^{\prime}$ such that  
$\phi_q \neq 0$, $\phi_{cl}=\lambda D/2$.

Rearranging  Eq \ref{dysona1} explicitly  we find that $G^K$ and $G^Z$ obey the equations
\begin{eqnarray}
G^Z &=& \frac{\bar{g}^A \phi_{gq}}{1-{\bar g}^R\phi_{g q}{\bar g}^A\phi_{g q}}\left[\bar{g}^R + \bar{g}^K( {\bar g}^{A})^{-1} G^Z \right]
\label{grq}\\
G^K &=& \frac{1}{1-{\bar g}^R\phi_{g q}{\bar g}^A\phi_{g q}} \left[\bar{g}^K + \bar{g}^R \phi_{gq} \bar{g}^A
+ \bar{g}_K \phi_{gq} G^K \right]
\label{gkq}
\end{eqnarray}
Eq~\ref{grq} and Eq~\ref{gkq} are singular integral equations.
Noting that $G^Z(t,t') \neq 0$  only when $\phi_q(t)\neq 0$ and that ${\bar g}^{R,A}$ 
are effectively delta functions, we see that the long time behaviors of $G^K$ and $G^Z$ are the same. 
In equilibrium we may set $\mu_{L,R}=0$ and write Eq \ref{gkq} explicitly for $t,t^\prime$
separated widely in time. The important term is the last one, which is 
\begin{equation}
\int^\prime dt_1 \frac{1}{t-t_1}\frac{\lambda}{1-\frac{\lambda^2}{4}}G^K(t_1,t^{\prime})
\label{GKEQ}
\end{equation}
where the prime denotes an integration only over those times $t_1$ for which $\phi_q(t_1)\neq 0$.
From the standard properties of singular integral equations \cite{Nozieres69,Anderson69} we identify the
coefficient as the tangent of the  phase
shift, obtaining 
\begin{equation}
\tan \delta_{eq}=\frac{\lambda}{1-\frac{\lambda^2}{4}}
\label{deltaeq}
\end{equation}
and recovering the usual equilibrium Coulomb gas.

In the non-equilibrium long time limit we follow Ng \cite{Ng96} and
write $G^K=e^{-i\mu_L(t-t^\prime)}G^K_L+e^{-i\mu_R(t-t^\prime)}G^K_R$.
We substitute this expression into Eq \ref{gkq}, write separate equations for $G^K_L$ and $G^K_R$,
use Eq~\ref{bargk} and note that for $\Delta \mu (t-t^\prime)>>1$ 
the cross term between the $e^{i\mu_L}$ term
in $g^K$ and the $e^{i\mu_R}$ term in $G^K_R$ gives an effective delta function contribution
to the equation for $G^K$. This leads to a singular term of the form of Eq \ref{GKEQ}
but with the phase shift replaced by
\begin{eqnarray}
\tan \delta_{a=L,R}(t) = \frac{2D\phi_{gq}(t)cos^2\theta_a }{D^2-\phi^2_{gq}(t)-isgn(\mu_L-\mu_R)sin^2\theta_a2\phi_{gq}(t) D}
\label{phaseaa}
\end{eqnarray}
Note from Eq.~\ref{phaseaa} that $\delta_a(-\phi_{g q})= -\delta^*_a(\phi_{g q})$.

We briefly discuss the structure of the solution in the long-time nonequilibrium limit. 
From Eq.~\ref{closedloop}, we are eventually interested in the equal time limit of the Green's functions, which just as in equilibrium,
have a divergent contribution due to the long time approximation made in deriving
them. This issue
can be resolved by solving for $G_{K,Z}$ by assuming that the Green's functions adiabatically follow the 
time dependent potential. The non-divergent part of 
$G_{K,Z}$ leads to the logarithmic interaction between the charges, which has the following form
\begin{eqnarray}
C_0^{ln} = -\int_0^1 \frac{d g}{g} \int dt \sum_{a=L,R}
\left[\frac{\phi_{g q}}{\pi}
\frac{\partial \delta_a}{\partial \phi_{g q}}\right](t) \frac{\partial \ln{X_a(t)}}{\partial t}
\label{Cln} 
\end{eqnarray}
where 
\begin{eqnarray}
X_{a=L,R}(t) = \exp \left[\frac{P}{2\pi i} \int \frac{dt^{\prime \prime}}{t-t^{\prime \prime}} 
\ln \frac{1 - i \tan \delta_{a}(t^{\prime \prime})}{1+ \tan \delta_a(t^{\prime \prime})}\right] 
\label{X}
\end{eqnarray}

Using Eq.~\ref{X}, one may rewrite $\ln X_a(t) = -\frac{1}{\pi}
P \int dt^{\prime \prime}\frac{\delta_a(t^{\prime \prime})}{(t-t^{\prime \prime})}$, further integrating Eq.~\ref{Cln}
by parts one finds
\begin{eqnarray}
C_0^{ln} 
= -\frac{1}{\pi^2}P\int_0^1 \frac{dg}{g} \int dt dt^{\prime \prime}\sum_{a=L,R}
\left[\phi_{g q} \frac{\partial \delta_a}{\partial \phi_{g q}}\right](t)
\frac{1}{t^{\prime \prime}-t} \frac{\delta_a(t^{\prime \prime})}{dt^{\prime \prime}} \label{cln1}
\end{eqnarray}
Our model allows for a sequence of quantum fields which alternates between $0$ and $\pm \frac{g\lambda D}{2}$, 
and therefore
may be written as,
\begin{equation}
\phi_{g q}(t) = \frac{g\lambda D}{2} \sum_{i=1\ldots} q_i \left[\theta(t-t_{2i-1})-\theta(t-t_{2i})\right]
\end{equation}
$i$ denoting times at which the quantum field changes, while, $q_i$ denoting the sign of the quantum
field in the region where it is nonzero.
From Eq.~\ref{phaseaa} it follows that 
\begin{equation}
\frac{\delta_{a}(t)}{dt} = \sum_{i=1\ldots}\delta_{a}(q_i) \left[  \delta(t-t_{2i-1}) 
-\delta(t-t_{2i})\right] 
\label{stepa}
\end{equation}
In a similar manner 
\begin{equation}
\left[\phi_{g q} \frac{\partial \delta_a}{\partial \phi_{g q}}\right](t)
=  \sum_{i=1\ldots}\left[ \phi_{g q} \frac{\partial \delta_a}{\partial \phi_{g q}}
\right](q_i)
\left[\theta(t-t_{2i-1})-\theta(t-t_{2i})\right]
\label{stepb}
\end{equation}
Substituting Eq.~\ref{stepa} and ~\ref{stepb} in Eq.~\ref{cln1}, we find
\begin{eqnarray}
C^{ln}= -\frac{1}{\pi^2}P\int_0^1 \frac{dg}{g} \int dt \sum_{a=L,R;j} 
 \left[ \phi_{g q} \frac{\partial \delta_a}{\partial \phi_{g q}}
\right](q_j)
\left[\theta(t-t_{2j-1})-\theta(t-t_{2j})\right] \sum_i \delta_{a}(q_i) \left[\frac{1}{t_{2i-1}-t} 
-\frac{1}{t_{2i}-t}\right] 
\end{eqnarray}
The integral over time $t$ give rise to the logarithmic interaction between charges.
Moreover the coefficient of the logarithmic interaction between charges at
$t_i$ and $t_j$ depend on the quantum fields $q_{i,j}$. In particular if $q_i=q_j=q$,
the coupling constant integral yields a coefficient 
\begin{equation}
2\sum_{a=L,R} 
\int_0^1 \frac{dg}{g} \delta_a(q)\left[\phi_{g q} \frac{\partial \delta_a}
{\partial \phi_{g q}}\right](q)=\sum_{a=L,R}\delta_a^2 
\end{equation}
On the other hand, if 
$q_i=-q_j=q$, the coefficient of the logarithm interaction between charges is
\begin{equation}
\int_0^1 \frac{dg}{g} \sum_{a=L,R} \left(\delta^*_a(q)
\left[\phi_{g q}\frac{\partial \delta_a}
{\partial \phi_{gq}}\right](q)+ \delta_a(q) \left[\phi_{g}\frac{\partial \delta_a}
{\partial \phi_{gq}}\right]^*(q)  \right)
= \sum_{a=L,R}\delta_a \delta^*_a
\end{equation}
The overall $\pm$ signs before the coefficients of the logarithmic interaction
essentially keep track of whether the spin has flipped up or down and may be used to define
the Coulomb gas charge $n_i$.

The discussion so far is valid for $\Delta\mu t \gg 1$.
In order to obtain an expression for $C_0$  for 
arbitrary $\Delta \mu t$, we solve the Dyson equation
perturbatively, and obtain expressions for $C_0$ correct to second
order in the scattering potential $\lambda$. This is outlined in
Section \ref{hfunc}. The expression obtained interpolates the exact analytic
expressions for $\Delta \mu t \ll 1$ and $\Delta \mu t \gg 1$.

\section{Perturbative evaluation of $C_0$ for a symmetric and hard cutoff}
\label{hfunc}

Let us turn to the evaluation of the time
evolution operator at $O(\Delta_T^2)$ (which 
corresponds to a single instanton in the quantum field),
\begin{equation}
e^{-C_0(t)} = Tr_{el} \left[T e^{-i \int _0^t
dt^{\prime}\tilde{V}_{\uparrow}(t^{\prime})}
\rho_{S0}^{\uparrow}\right]
\end{equation}
with $\tilde{V}_{\uparrow}(t^{\prime})$ defined in Eq.~\ref{scatpot}.

By using the convenient 
representation 
$d_{\alpha}(t) = \frac{1}{\sqrt{\pi}} \sum_{k,a=L,R}\cos\theta_a c_{k \alpha a}(t)$, 
the expression for $C_0$ at $O(\lambda^2)$ is
\begin{equation}
C_0(t) = \frac{N\lambda^2}{\pi^2}\sum_{\alpha \beta=L,R}
 \cos^2\theta_{\alpha} \cos^2\theta_{\beta} \int_{0}^{t} dt_1
\int_0^{t_1} dt_2 \int_{-\xi^{-1} + \mu_{\alpha}}^{\mu_{\alpha}}
d\epsilon_1 \int_{\mu_{\beta}}^{{\xi}^{-1}+\mu_{\beta}}
d\epsilon_2\langle c^{\dagger}_{\epsilon_1 \alpha}(t_1)
c_{\epsilon_1\alpha}(t_2) \rangle \langle c_{\epsilon_2 \beta}(t_1)
c^{\dagger}_{\epsilon_2 \beta}(t_2) \rangle
\end{equation}
where $0\leq \theta_L = \frac{\pi}{2}-\theta_R\leq \frac{\pi}{2}$.
The limits of integration for $\epsilon_{1,2}$ correspond to band
edges that are abrupt and symmetrically located with respect to the
chemical potentials (c.f. Fig~\ref{VD}). At zero temperatures,
\begin{eqnarray}
C_0(t)&=& \frac{N\lambda^2}{\pi^2}\sum_{\alpha \beta=L,R}
\cos^2\theta_{\alpha} \cos^2\theta_{\beta}\int_{0}^{t} dt_1
\int_0^{t_1} dt_2  \int_{-\xi^{-1} + \mu_{\alpha}}^{\mu_{\alpha}}
d\epsilon_1 \int_{\mu_{\beta}}^{{\xi}^{-1}+\mu_{\beta}} d\epsilon_2
e^{-i(\epsilon_1 - \epsilon_2)(t_2 - t_1)} \nonumber \\
&=& -\frac{N\lambda^2}{\pi^2}\sum_{\alpha \beta=L,R}
\cos^2\theta_{\alpha} \cos^2\theta_{\beta} \int_{0}^{t} dt_1
\int_0^{t_1} dt_2 e^{-i(\mu_{\beta}-\mu_{\alpha})(t_1 -
t_2)}\frac{\left[1 - e^{-i(t_1 - t_2)/\xi} \right]^2}{(t_1 - t_2)^2}\label{b3}
\end{eqnarray}
It is also of interest to define a soft cutoff model with density of states
$\rho_{L,R}(\epsilon) \sim e^{- |\epsilon - \mu_{L,R}|\xi}$. This gives Eq.~\ref{b3}
but with $\frac{\left[1 - e^{-i (t_1 - t_2)/\xi} \right]}{t_1 - t_2} \rightarrow
\frac{1}{t_1 - t_2 - i \xi}$.

Writing the above as a symmetric and anti-symmetric combination of
the time arguments $t_1,t_2$ one obtains
\begin{equation}
C_0 = Re[C_0] + Im[C_0]
\end{equation}
where $Re[C_0]$ is given by the symmetric combination of $t_1,t_2$
which after a straightforward evaluation of time integrals leads to,
\begin{eqnarray}
Re[C_0](t,\Delta \mu,\xi)&=&
N\lambda^2(\frac{\cos^4\theta_L}{\pi^2}+\frac{\cos^4\theta_R}{\pi^2})\left[F_0(2t/\xi)-2F_0(t/\xi)
\right] \nonumber \\
&+& 2N\lambda^2\frac{\cos^2\theta_L \cos^2\theta_R}{\pi^2} \left[
F_0(\Delta \mu t) + \frac{F_0(2t/\xi + \Delta \mu t)+F_0(2t/\xi -
\Delta \mu t)}{2} - F_0(t/\xi + \Delta \mu t) -  F_0(t/\xi - \Delta
\mu t)\right]
\nonumber\\
&+&  2N\lambda^2\frac{\cos^2\theta_L \cos^2\theta_R}{\pi^2}\left[
F_1(\Delta \mu t) + \frac{F_1(2t/\xi + \Delta \mu t)+F_1(2t/\xi -
\Delta \mu t)}{2} - F_1(t/\xi + \Delta \mu t) - F_1(t/\xi - \Delta
\mu t)\right]\nonumber
\end{eqnarray}
with
\begin{eqnarray}
F_0(x) &=& -\ln|x| - \gamma + Ci(x) + (\cos x -1)\label{f0}\\
F_1(x) &=& x Si(x)\label{f1}
\end{eqnarray}
where $Ci(z) = \gamma + \ln z + \int_0^{z}dt \frac{\cos t -1}{t}$.
Note $F_0(x\ll 1) = -x^2/2$, $F_1(x\ll 1) = x^2$, $F_1(x\gg 1) =
\frac{\pi}{2}x$.

The antisymmetric combination of $t_1,t_2$ leads to
\begin{eqnarray}
Im[C_0(t)] \propto i t
\end{eqnarray}
and represents unimportant energy renormalization that vanishes for the particle-hole symmetric case.

Identifying the coefficients above with perturbative expressions for
the appropriate phase shifts defined in Eq.~\ref{phaseshift}, and
defining the functions,
\begin{eqnarray}
h_0(t/\xi) &=& \left[F_0(2t/\xi)-2F_0(t/\xi)\right]
\\
h_M(t/\xi,\Delta \mu t) &=& \left[F_0(\Delta \mu t) +
\frac{F_0(2t/\xi + \Delta \mu t)+F_0(2t/\xi - \Delta \mu t)}{2} -
F_0(t/\xi + \Delta \mu t) - F_0(t/\xi - \Delta \mu t) \right]
\\h_{neq}(t/\xi,\Delta \mu
t)&=& \frac{2}{\pi}\left[F_1(\Delta \mu t) + \frac{F_1(2t/\xi +
\Delta \mu t)+F_1(2t/\xi - \Delta \mu t)}{2} - F_1(t/\xi + \Delta
\mu t) - F_1(t/\xi - \Delta \mu t)\right]
\end{eqnarray}
one obtains Eq.~\ref{C}.

To obtain the scaling function $g_{neq}$ giving rise to the additive renormalization
of $\Gamma_{neq}$, we differentiate Eq.~\ref{b3} in its soft cutoff analogue with respect to
$\ln \xi$, and (to extract the long time behavior) $t$. The resulting integral may
easily be evaluated. We plot the real part in Fig~\ref{crossover}.

\end{document}